\documentclass[11pt]{article}

\usepackage{url}
\usepackage{amsmath}

\usepackage{sgame}
\usepackage{color}

\def\smallromani{\renewcommand{\theenumi}{\roman{enumi}}
\renewcommand{\labelenumi}{(\theenumi)}}

\newcommand{\Proof}{\NI
                    {\bf Proof.}\ }
\newtheorem{theorem}{Theorem}[section]
\newtheorem{defined}[theorem]{Definition}
\newenvironment{definition}{\begin{defined} \rm}{\end{defined}}
\newtheorem{exa}[theorem]{Example}
\newenvironment{example}{\begin{exa} \rm}{\end{exa}}

\newtheorem{lemma}[theorem]{Lemma}
\newtheorem{corollary}[theorem]{Corollary}

\newtheorem{note}[theorem]{Note}

\newtheorem{claim}{Claim}
\newtheorem{exe}{Exercise}

\newtheorem{pro}{Problem}

\newcounter{symbol}
\newcommand{\indexsyma}[1]%
{\stepcounter{symbol}\index{zzz1 \thesymbol @\protect#1}}
\newcommand{\indexsymb}[1]%
{\stepcounter{symbol}\index{zzz2 \thesymbol @\protect#1}}
\newcommand{\indexsymc}[1]%
{\stepcounter{symbol}\index{zzz3 \thesymbol @\protect#1}}
\newcommand{\indexsymd}[1]%
{\stepcounter{symbol}\index{zzz4 \thesymbol @\protect#1}}
\newcommand{\indexsyme}[1]%
{\stepcounter{symbol}\index{zzz5 \thesymbol @\protect#1}}

\newcommand{\oldbfe}[1]{\begin{bfseries}\emph{#1}\end{bfseries}}

\newcommand{\ES}{\mbox{$\emptyset$}}

\newcommand{\myra}{\mbox{$\:\rightarrow\:$}}

\newcommand{\Ra}{\mbox{$\:\Rightarrow\:$}}

\newcommand{\sse}{\mbox{$\:\subseteq\:$}}

\newcommand{\fa}{\mbox{$\forall$}}
\newcommand{\te}{\mbox{$\exists$}}

\newcommand{\LL}{\mbox{$\ldots$}}

\newcommand{\C}[1]{\mbox{$\{{#1}\}$}}           

\newcommand{\NI}{\noindent}
\newcommand{\HB}{\hfill{$\Box$}}
\newcommand{\VV}{\vspace{5 mm}}

\newcommand{\II}{\vspace{2 mm}}

\newcommand{\szkew}[1]{\relax \setbox0=\hbox{\kern -24pt $\displaystyle#1$\kern 0pt }%
\box0}
{\catcode`\@=11 \global\let\ifjusthvtest@=\iffalse}

\newcounter{oldmycaption}

\usepackage{handbookbib}
\usepackage{amsfonts}
\usepackage{latexsym}
\usepackage{alltt}

\title{Order Independence and Rationalizability}

\author{Krzysztof R. Apt \\
\emph{School of Computing, National University of Singapore} \\
\emph{3 Science Drive 2, Republic of Singapore 117543}
\footnote{On leave from CWI, Amsterdam, the Netherlands and
University of Amsterdam.}
}
\begin{document}

\date{}

\maketitle

\pagestyle{empty}

\thispagestyle{empty}

\begin{abstract}
  Two natural strategy elimination procedures have been studied for
  strategic games. The first one involves the notion of (strict, weak,
  etc) dominance and the second the notion of rationalizability. In
  the case of dominance the criterion of order independence allowed us
  to clarify which notions and under what circumstances are robust.
  In the case of rationalizability this criterion has not been
  considered.
  
  In this paper we investigate the problem of order independence for
  rationalizability by focusing on three naturally entailed reduction
  relations on games. These reduction relations are distinguished by
  the adopted reference point for the notion of a better response.
  Additionally, they are parametrized by the adopted system of
  beliefs.
  
  We show that for one reduction relation the outcome of its (possibly
  transfinite) iterations does not depend on the order of elimination
  of the strategies.  This result does not hold for the other two
  reduction relations.  However, under a natural assumption the
  iterations of all three reduction relations yield the same outcome.
  
  The obtained order independence results apply to the frameworks
  considered in \cite{Ber84} and \cite{Pea84}.  For finite games the
  iterations of all three reduction relations coincide and the order
  independence holds for three natural systems of beliefs considered
  in the literature.
\end{abstract}

\section{Introduction}

Rationalizability was introduced in \cite{Ber84} and \cite{Pea84} to
formalize the intuition that players in non-cooperatives games act by
having common knowledge of each others' rational behaviour.
Rationalizable strategies in a strategic game are defined as a limit
of an iterative process in which one repeatedly removes the strategies
that are never best responses (NBR) to the beliefs held about the other
players.  In contrast to the iterated elimination of strictly and of
weakly dominated strategies at each stage all `undesirable' strategies
are removed.

Much attention was devoted in the literature to the issue of order
independence for the iterated elimination of strictly and of weakly
dominated strategies. It is well-known that strict dominance is order
independent for finite games (see \cite{GKZ90} and \cite{Ste90}),
while weak dominance is order dependent.  This has been often used as
an argument in support of the first procedure and against the second
one, see, e.g., \cite{OR94}.  On the other hand, \cite{DS02} indicated
that order independence for strict dominance fails for
arbitrary games though does hold for a large class of infinite games.

The criterion of order independence did not seem to be applied to
assess the merits of the iterated elimination of NBR.  In this paper
we study this problem by analyzing what happens when at each stage of
the iterative process only some strategies that are NBR are
eliminated.  This brings us to a study of three naturally entailed
reduction relations.  They are distinguished by the adopted reference
point for the notion of a better response, which can be the initial
game, the game currently being reduced or the reduced game.
Additionally, they are parametrized by the adopted system of beliefs.
In general these relations differ and transfinite iterations are possible.

We show for one reduction relation that for all `well-behaving'
systems of beliefs the outcome of the iterated elimination of strategies
does not depend on the order of elimination.  The result does not
hold for the other two reduction relations, even for two-person games
and beliefs being the strategies of the opponent.

Further, using a game modeling a version of Bertrand competition between two
firms we show that the variants of these reduction relations in
which all strategies that are NBR are eliminated differ, as well. The
same example also shows that the relation considered in \cite{Ber84},
according to which at each stage \emph{all} strategies that are NBR
are eliminated, yields a weaker reduction than the one according to
which at each stage only \emph{some} strategies that are NBR are
eliminated. In other words, natural games exist in which it is beneficial to
eliminate at certain stages only some strategies that are NBR.

The situation changes if we assume that for each belief $\mu_i$ in a
restriction $G$ of the original game a best response to $\mu_i$ in $G$
exists.  We show that then the iterations of all three reduction
relations yield the same outcome.  This implies order independence for
all three reduction relations for the class of games for which
\cite{DS02} established order independence of the iterated elimination
of strictly dominated strategies.

A complicating factor in these considerations is that iterations of
each of the reduction relation can reduce the initial game to an empty
game. We discuss natural examples of games for which the unique
outcome of the iterated elimination process is a non-empty game.
In particular, order independence and non-emptiness of the
final outcome holds for a relaxation of two elimination procedures studied
in the literature:

\begin{itemize}
\item the one considered in \cite{Ber84}, concerning a compact game
  with continuous payoff functions, in which at each stage we now eliminate 
  only \emph{some} strategies that are NBR (to the joint strategies of the
  opponents), and
  
\item the one considered in \cite{Pea84}, concerning mixed extension
  of a finite game, in which at each stage we now eliminate only
  \emph{some} mixed strategies that are NBR (to the elements of the
  products of convex hulls of the opponents' strategies).
\end{itemize}

The definition of rationalizable strategies is parameterized by a
system of belief.  In the case of finite games three
natural alternatives were considered:

\begin{itemize}
\item joint pure strategies of the opponents,
see, e.g., \cite{Ber84},

\item joint mixed strategies of the opponents,
see, e.g., \cite{Ber84} and \cite{Pea84},

\item probability distributions over the
joint pure strategies of the opponents, see, e.g., \cite{Ber84} and \cite{OR94}.
\end{itemize}
A direct consequence of our results is that for finite games order
independence holds for all three reduction relations and all three
alternatives of the systems of belief.

In summary, all three versions of the iterated elimination of NBR
are order independent for the same classes of games for
which iterated elimination of strictly dominated strategies was
established.  Additionally, for one version order independence
holds for all `well-behaving' systems of beliefs.

\section{Preliminaries}

Given $n$ players we represent a
strategic game (in short, a game)
by a sequence
\[
(S_1, \LL, S_n, p_1, \LL, p_n),
\] 
where for each $i \in [1..n]$

\begin{itemize}
\item $S_i$ is the non-empty set of \oldbfe{strategies} 
available to player $i$,

\item $p_i$ is the payoff function for the  player $i$, so
$
p_i : S_1 \times \LL \times S_n \myra \cal{R},
$
where $\cal{R}$ is the set of real numbers.
\end{itemize}

Given a sequence of non-empty sets of strategies $S_1, \LL, S_n$ and
$s \in S_1 \times \LL \times S_n$ we denote the $i$th element of $s$ by $s_i$ and
use the following standard notation:

\begin{itemize}
\item $s_{-i} := (s_1, \LL, s_{i-1}, s_{i+1}, \LL, s_n)$,

\item $(s'_i, s_{-i}) := (s_1, \LL, s_{i-1}, s'_i, s_{i+1}, \LL, s_n)$, where
we assume that $s'_i \in S_i$,

\item $S_{-i} := S_1 \times \LL \times S_{i-1} \times S_{i+1} \times \LL \times S_n$.

\end{itemize}
We denote the strategies of player $i$ by $s_i$, possibly with some superscripts.

By a \oldbfe{restriction} of a game $H := (T_1, \LL, T_n, p_1,
\LL,p_n)$ we mean a game $G := (S_1, \LL, S_n, p_1,$  $\LL, p_n)$ such
that each $S_i$ is a (possibly empty) subset of $T_i$ and each $p_i$ is
identified with its restriction to the smaller domain. 
We write then $G \sse H$.

If some $S_i$ is empty, we call $G$ a \oldbfe{degenerate restriction} of $H$.  In
this case the references to $p_j(s)$ (for any $j \in [1..n]$)
are incorrect and we shall need to be careful about this.
If all $S_i$ are empty, we call $G$ an \oldbfe{empty game} and denote it by $\ES_n$.
If no $S_i$ is empty, we call $G$ a \oldbfe{non-degenerate restriction} of $H$.

Similarly, we introduce the notions of a \oldbfe{union}
and \oldbfe{intersection} of a transfinite sequence
$(G_{\alpha})_{\alpha < \gamma}$ of restrictions of $H$ ($\alpha$ and
$\gamma$ are ordinals) denoted respectively by $\bigcup_{\alpha <
  \gamma} G_{\alpha}$ and $\bigcap_{\alpha < \gamma} G_{\alpha}$.

\section{Belief structures}

We assume that each player $i$ in the game $H = (T_1, \LL, T_n, p_1,
\LL,p_n)$ has some further unspecified non-empty set of beliefs ${\cal
  B}_{i}$ about his opponents.  We call then ${\cal B} := ({\cal
  B}_{1}, \LL, {\cal B}_{n})$, a \oldbfe{belief system} in the game
$H$.  We further assume that each payoff function $p_i$ can be
modified to an \oldbfe{expected payoff} function $p_i : S_i \times
{\cal B}_{i} \myra \cal{R}$.

Then we say that a strategy $s_i$ of player $i$ is a \oldbfe{best
  response} to a belief $\mu_i \in {\cal B}_{i}$ in $H$ if for all
strategies $s'_i \in T_i$
\[
p_i(s_i, \mu_i) \geq p_i(s'_i, \mu_i).
\]

In what follows we also assume that each set of beliefs ${\cal B}_{i}$
of player $i$ in $H$ can be \emph{narrowed} to any restriction $G$ of
$H$. We denote the outcome of this \oldbfe{narrowing} of ${\cal
  B}_{i}$ to $G$ by ${\cal B}_{i} \stackrel{.}{\cap} G$.  The beliefs
in ${\cal B}_{i} \stackrel{.}{\cap} G$ can be also considered as
beliefs in the game $G$.  We call then the pair $({\cal B},
\stackrel{.}{\cap})$, where ${\cal B} := ({\cal B}_{1}, \LL, {\cal
  B}_{n})$, a \oldbfe{belief structure} in the game $H$.

Finally, given a belief structure $({\cal B}, \stackrel{.}{\cap})$ in a game $H$ we say
that a restriction $G$ of $H$ is ${\cal B}$-\oldbfe{closed} if each
strategy $s_i$ of player $i$ in $G$ is a best response in $H$ (note
this reference to $H$ and \emph{not} $G$) to a belief in ${\cal B}_{i}
\stackrel{.}{\cap} G$.  

Fix now a game $H := (T_1, \LL, T_n, p_1, \LL, p_n)$ and a belief
structure $({\cal B}, \stackrel{.}{\cap})$ in $H$.  The following
natural property of $\stackrel{.}{\cap}$ will be relevant.

\begin{description}
\item[A] If $G _1\sse G_2 \sse H$, then for all $i \in [1..n]$, 
${\cal B}_{i} \stackrel{.}{\cap} G_1 \sse {\cal B}_{i} \stackrel{.}{\cap} G_2$.
\end{description}

The following belief structure will be often used.  Assume
that for each player his set of beliefs ${\cal B}_{i}$ in the game $H
:= (T_1, \LL, T_n, p_1, \LL, p_n)$ consists of the joint strategies of
the opponents, i.e., ${\cal B}_{i} = T_{-i}$.  For a restriction $G :=
(S_1, \LL, S_n, p_1, \LL, p_n)$ of $H$ we define $T_{-i}
\stackrel{.}{\cap} G := S_{-i}$.  
Note that property \textbf{A} is then satisfied.  
We call $({\cal B}, \stackrel{.}{\cap})$ the \oldbfe{pure}
\footnote{to indicate that it involves only pure strategies of the opponents}
belief structure in $H$.

\section{Reductions of games}
\label{sec:reductions}

Assume now a game $H$ and a belief structure $({\cal B}, \stackrel{.}{\cap})$ in $H$.
We introduce a notion of reduction $\leadsto$ between a
restriction $G := (S_1, \LL, S_n, p_1, \LL, p_n)$ of $H$ and a
restriction $G' := (S'_1, \LL, S'_n, p_1, \LL, p_n)$ of $G$ defined by:

\begin{itemize}
\item $G \leadsto G'$ when $G \neq G'$ and for all $i \in [1..n]$ 
\[
\mbox{no $s_i \in S_i \setminus S'_i$ is a best response in $H$ to some $\mu_{i} \in {\cal B}_{i} \stackrel{.}{\cap} G$.}
\]
\end{itemize}

Of course, the $\leadsto$ relation depends on the underlying belief
structure $({\cal B}, \stackrel{.}{\cap})$ in $H$ but we do not
indicate this dependence as no confusion will arise.  Note that in the
definition of $\leadsto$ we do not require that all strategies that
are NBR are removed. So in general $G \leadsto G'$ can hold for
several restrictions $G'$.  Also, what is important, we refer to the
best responses in $H$ and \emph{not} in $G$ or $G'$.  The reduction
relations that take these two alternative points of reference will be
studied in the next section.

Let us define now appropriate iterations of the $\leadsto$ relation.
We shall use this concept for various reduction relations so define it for an arbitrary relation
$\longmapsto$ between a restriction $G$ of $H$ and a restriction $G'$ of $G$.

\begin{definition}
Consider a transfinite sequence of restrictions $(G_{\alpha})_{\alpha \leq \gamma}$ of $H$ such that
  \begin{itemize}

  \item $H = G_{0}$,

  \item for all $\alpha < \gamma$, $G_{\alpha} \longmapsto G_{\alpha + 1}$,

  \item for all limit ordinals $\beta \leq \gamma$, $G_{\beta} = \bigcap_{\alpha < \beta} G_{\alpha}$,

  \item for no $G'$, $G_{\gamma} \longmapsto G'$ holds.
  \end{itemize}
We say then that $(G_{\alpha})_{\alpha \leq \gamma}$ is a \oldbfe{maximal sequence} of the $\longmapsto$ reductions and call  $G_{\gamma}$ its \oldbfe{outcome}. Also, we write  $H \longmapsto^{\alpha} G_{\alpha}$ for each $\alpha \leq \gamma$.
\HB  
\end{definition}

We now establish the following general order independence result. 

\begin{theorem}[Order Independence] \label{thm:order}
  Consider a game $H$ and a belief structure $({\cal B},
  \stackrel{.}{\cap})$ in $H$.  Assume property \textbf{A}.  Then any
  maximal sequence of the $\leadsto$ reductions yields the
  same outcome which is the largest restriction of $H$ that is ${\cal
    B}$-closed.

\end{theorem}
\Proof 
First we establish the following claim.

\begin{claim} \label{cla:largest}
There exists a largest restriction of $H$ that is ${\cal B}$-closed.
\end{claim}
\emph{Proof}.
First note that each empty game is ${\cal B}$-closed.
Consider now a transfinite sequence of restrictions
$(G_{\alpha})_{\alpha < \gamma}$ of $H$ such that each $G_{\alpha}$ is
${\cal B}$-closed.  We claim that then $\bigcup_{\alpha < \gamma}
G_{\alpha}$ is ${\cal B}$-closed, as well.

To see this choose a strategy $s_i$ of player $i$ in $\bigcup_{\alpha < \gamma} G_{\alpha}$.
Then $s_i$ is a strategy of player $i$ in $G_{\alpha_{0}}$ for some $\alpha_{0} < \gamma$. 
The restriction $G_{\alpha_{0}}$ is ${\cal B}$-closed, so
for some $\mu_i \in {\cal B}_{i} \stackrel{.}{\cap} G_{\alpha_{0}}$ the strategy 
$s_i$ is a best response to $\mu_i$ in $H$.
By property \textbf{A} $\mu_i \in {\cal B}_{i} \stackrel{.}{\cap} \bigcup_{\alpha < \gamma} G_{\alpha}$.
\HB
\II

Consider now a maximal sequence
  $(G_{\alpha})_{\alpha \leq \gamma}$ of the $\leadsto$ reductions. 
Take a  restriction $H'$ of $H$ such that for some $\alpha < \gamma$
\begin{itemize}

\item $H'$ is  ${\cal B}$-closed,

\item $H' \sse G_{\alpha}$.

\end{itemize}
Consider a strategy $s_i$ of player $i$ in $H'$. Then $s_i$ is also a strategy of player $i$ in $G_{\alpha}$.
$H'$ is ${\cal B}$-closed, so 
$s_i$ is a best response in $H$
to a belief $\mu_i \in {\cal B}_{i} \stackrel{.}{\cap} H'$.
By property \textbf{A} $\mu_i \in {\cal B}_{i} \stackrel{.}{\cap} G_{\alpha}$.
So by the definition of the $\leadsto$ reduction the strategy $s_i$ is not
deleted in the transition $G_{\alpha} \leadsto G_{\alpha + 1}$, i.e., 
$s_i$ is a strategy of player $i$ in $G_{\alpha +1}$.
Hence $H' \sse G_{\alpha + 1}$. 

We conclude by transfinite induction that $H' \sse G_{\gamma}$.
In particular we conclude that
$G_{\cal B} \sse G_{\gamma}$,
where $G_{\cal B}$ is the largest restriction of $H$ that is ${\cal B}$-closed
and the existence of which is guaranteed by Claim \ref{cla:largest}.

But also $G_{\gamma} \sse G_{\cal B}$ since $G_{\gamma}$ is ${\cal
  B}$-closed and $G_{\cal B}$ is the largest restriction of $H$ that
is ${\cal B}$-closed.

\HB
\VV

Since at each stage of the above elimination process some strategy is
removed, this iterated elimination process eventually stops, i.e., the
considered maximal sequences always exist.  The result can be
interpreted as a statement that each, possibly transfinite, iterated
elimination of NBR yields the same outcome.

The $\leadsto$ reduction 
allows us to remove only \emph{some} strategies that are NBR in the initial game $H$.
If we remove \emph{all} strategies that are NBR, we get the reduction relation 
that corresponds to the ones
considered in the literature for specific belief structures.
It is defined as follows.
Consider a restriction $G := (S_1, \LL, S_n, p_1, \LL, p_n)$ of $H$ and a
restriction $G' := (S'_1, \LL, S'_n, p_1, \LL, p_n)$ of $G$.
We define then the `fast' reduction $^{f} \hspace{-1mm} \leadsto$ by:

\begin{itemize}
\item $G \hspace{1mm} ^{f} \hspace{-1mm} \leadsto G'$ when $G \neq G'$ and for all $i \in [1..n]$ 
\[
S'_i = \{ s_i \in S_i \mid \te \mu_{i} \in {\cal B}_{i} \stackrel{.}{\cap} G \: \fa s'_i \in T_i \:
p_{i}(s'_i, \mu_{i}) \leq p_{i}(s_i, \mu_{i})\}.
\]
\end{itemize}

Since the $^{f} \hspace{-1mm} \leadsto$ reduction removes all
strategies that are never best responses, $G \hspace{1mm} ^{f} \hspace{-1mm} \leadsto G'$
and $G \leadsto G''$ implies $G' \sse G''$.

We now show that the iterated application of the $^{f} \hspace{-1mm}
\leadsto$ reduction yields a stronger reduction than 
$\leadsto$ and that $^{f} \hspace{-1mm} \leadsto$ is indeed `fast' in
the sense that it generates reductions of the original game $H$ faster
than the $\leadsto$ reduction.
While this is of course as expected, we shall see in the next section that these
properties do not hold for a simple variant of the $\leadsto$ reduction
studied in the literature.

\begin{theorem} \label{thm:fast}
Consider a game $H$ and a belief
structure $({\cal B}, \stackrel{.}{\cap})$ in $H$.
Assume property \textbf{A}. 

\begin{enumerate} \smallromani
  \item 
Suppose $G \hspace{1mm} ^{f} \hspace{-1mm} \leadsto^{\gamma} G'$
and $G \leadsto^{\gamma} G''$. Then $G' \sse G''$.

  \item 
Suppose $H \hspace{1mm} ^{f} \hspace{-1mm} \leadsto^{\beta} G$
and $H \leadsto^{\gamma} G$.
Then $\beta \leq \gamma$.
\end{enumerate} 
\end{theorem}

\Proof
First we establish a simple claim concerning the restrictions of $H$.

\setcounter{claim}{0}

\begin{claim} \label{claim1}
Suppose $G_1 \sse G_2$, 
$G_1 \hspace{1mm} ^{f} \hspace{-1mm} \leadsto G'$
and $G_2 \hspace{1mm} ^{f} \hspace{-1mm} \leadsto G''$.
Then $G' \sse G''$.
\end{claim}
\emph{Proof}.
Let $G' := (S'_1, \LL, S'_n, p_1, \LL, p_n)$ and
$G'' := (S''_1, \LL, S''_n, p_1, \LL, p_n)$.

Suppose $s'_i \in S'_i$. Then for some $\mu_{i} \in {\cal B}_{i}
\stackrel{.}{\cap} G_1$ we have $\fa s^{*}_i \in T_i \: p_{i}(s^{*}_i, \mu_{i})
\leq p_{i}(s'_i, \mu_{i})$.
By property \textbf{A} $\mu_{i} \in {\cal B}_{i} \stackrel{.}{\cap} G_2$, so
$s'_i \in S''_i$. 
\HB
\II

\NI
$(i)$
By definition appropriate transfinite sequences $(G'_{\alpha})_{\alpha \leq \gamma}$ 
and $(G''_{\alpha})_{\alpha \leq \gamma}$ such that $G = G'_{0} = G''_{0}$,
$G' = G'_{\gamma}$ and $G'' = G''_{\gamma}$ exist.
We proceed by transfinite induction.

Suppose the claim holds for all $\beta < \gamma$.
\II

\NI
\emph{Case 1}. $\gamma$ is a successor ordinal, say $\gamma = \beta +1$.

By the induction hypothesis $G'_{\beta} \sse G''_{\beta}$.
By Claim \ref{claim1} $G'_{\gamma} \sse G_2$, where 
$G''_{\beta} \hspace{1mm} ^{f} \hspace{-1mm} \leadsto G_2$.
But by the definition of the $^{f} \hspace{-1mm} \leadsto$ reduction
also $G_2 \sse G''_{\gamma}$. So $G'_{\gamma} \sse G''_{\gamma}$.
\II

\NI
\emph{Case 2}. $\gamma$ is a limit ordinal.

By the induction hypothesis for all $\beta < \gamma$ we have
$G'_{\beta} \sse G''_{\beta}$. By definition
$G'_{\gamma} = \bigcap_{\beta < \gamma} G'_{\beta}$ and
$G''_{\gamma} = \bigcap_{\beta < \gamma} G''_{\beta}$, so
$G'_{\gamma} \sse G''_{\gamma}$.
\II

\NI
$(ii)$
Let $(G_{\alpha})_{\alpha \leq \beta}$ and $(G'_{\alpha})_{\alpha \leq \gamma}$ 
be the sequences of the reduction of $H$ that respectively ensure
$H \hspace{1mm} ^{f} \hspace{-1mm} \leadsto^{\beta} G$
and $H \leadsto^{\gamma} G$.

Suppose now that on the contrary $\gamma < \beta$.  Then $G_{\beta}
\subset G_{\gamma}$ by the definition of the $^{f} \hspace{-1mm}
\leadsto$ reduction.  By $(i)$ we also have $G_{\gamma} \sse
G'_{\gamma}$. Further, $G'_{\gamma} = G_{\beta}$ since by assumption
both of them equal $G$, so $G_{\beta} = G_{\gamma}$, which is a
contradiction.  
\HB 
\VV

It is important to note that the outcome of the considered iterated elimination process
can be an empty game.

\begin{example} \label{exa:1}
Consider a two-players game $H$ in which the set of strategies for each player is
the set of natural numbers. The payoff to each player is the number (strategy) he selected.
Suppose that beliefs are the strategies of the opponent.
Clearly  no strategy is a best response
to a strategy of the opponent. So $H \leadsto \ES_2$.
\HB
\end{example}

In general, infinite sequences of the $\leadsto$ reductions
are possible. Even more, in some games $\omega$ steps of the
$^{f} \hspace{-1mm} \leadsto$ reduction are insufficient to reach a ${\cal
  B}$-closed game.

\begin{example} \label{exa:2}
Consider the following game $H$ with three players.
The set of strategies for each player is the set of natural numbers ${\cal N}$.
The payoff functions are defined as follows:
\[
p_1(k, \ell, m) := \left\{ 
\begin{tabular}{ll}
$k$ &  \mbox{if $k = \ell +1$} \\
0 &  \mbox{otherwise}
\end{tabular}
\right . 
\]
\[
p_2(k, \ell, m) := \left\{ 
\begin{tabular}{ll}
$k$ &  \mbox{if $k = \ell$} \\
0 &  \mbox{otherwise}
\end{tabular}
\right . 
\]
\[
p_3(k, \ell, m) := 0.
\]
  
Further we assume the pure belief structure.  
Each restriction of $H$ can be identified with the triple of the strategy sets of the players.
Note that

\begin{itemize}
\item the best response to $s_{-1} = (\ell, m)$ is $\ell +1$,

\item the best response to $s_{-2} = (k, m)$ is $k$,

\item each $m \in {\cal N}$ is a best response to $s_{-3} = (k, \ell)$.

\end{itemize}
So the following sequence of reductions holds:
\[
({\cal N}, {\cal N}, {\cal N}) \hspace{1mm} ^{f} \hspace{-1mm} \leadsto ({\cal N} \setminus \C{0}, {\cal N}, {\cal N}) \hspace{1mm} ^{f} \hspace{-1mm} \leadsto ({\cal N} \setminus \C{0}, {\cal N} \setminus \C{0}, {\cal N}) \hspace{1mm} ^{f} \hspace{-1mm} \leadsto 
\]
\[({\cal N} \setminus \C{0,1}, {\cal N} \setminus \C{0}, {\cal N}) \hspace{1mm} ^{f} \hspace{-1mm} \leadsto ({\cal N} \setminus \C{0,1}, {\cal N} \setminus \C{0,1}, {\cal N}) \hspace{1mm} ^{f} \hspace{-1mm} \leadsto \ \LL
\]
So $({\cal N}, {\cal N}, {\cal N}) \hspace{1mm} ^{f} \hspace{-1mm} \leadsto^{\omega} (\ES, \ES, {\cal N})$.
Also $(\ES, \ES, {\cal N}) \hspace{1mm} ^{f} \hspace{-1mm} \leadsto (\ES, \ES, \ES)$, so 
$({\cal N}, {\cal N}, {\cal N}) \hspace{1mm} ^{f} \hspace{-1mm} \leadsto^{\omega + 1} (\ES, \ES, \ES)$.

Further, it is easy to see that it is the only maximal sequence of the $\leadsto$
reductions.
\HB
\end{example}

Let us mention here that \cite{Lip94} constructed a two-player game
for which $\omega$ steps of the $^{f} \hspace{-1mm} \leadsto$ reduction
are not sufficient to reach a ${\cal
  B}$-closed game, where each ${\cal B}_i$ consists of the mixed
strategies of the opponent.

These examples bring us to the question: are we studying the right
reduction relation?

\section{Variations of the reduction relation}

Indeed, a careful reader may have noticed that we use a slightly
different notion of reduction than the one considered in \cite{Ber84}
and \cite{Pea84}. In general, two natural alternatives to the
$\leadsto$ relation exist. In this section we introduce
these variations and clarify when they coincide.

Given a restriction $G := (S_1, \LL, S_n, p_1, \LL, p_n)$ of a game $H
:= (T_1, \LL, T_n, p_1, \LL, p_n)$, a belief
structure $({\cal B}, \stackrel{.}{\cap})$ in $H$, where
${\cal B} := ({\cal B}_{1}, \LL, {\cal B}_{n})$ and a restriction $G' :=
(S'_1, \LL, S'_n, p_1, \LL, p_n)$ of $G$, the $\leadsto$
reduction can be alternatively defined by:

\begin{itemize}
\item $G \leadsto G'$ when $G \neq G'$ and for all $i \in [1..n]$ 
\[
\fa s_i \in S_i \setminus S'_i \: \fa \mu_{i} \in {\cal B}_{i} \stackrel{.}{\cap} G \: \te s'_i \in T_i \:
p_{i}(s'_i, \mu_{i}) > p_{i}(s_i, \mu_{i}).
\]
\end{itemize}
Two natural alternatives are:

\begin{itemize}
\item $G \myra G'$ when $G \neq G'$ and for all $i \in [1..n]$ 
\[
\fa s_i \in S_i \setminus S'_i \: \fa \mu_{i} \in {\cal B}_{i} \stackrel{.}{\cap} G \: \te s'_i \in S_i \:
p_{i}(s'_i, \mu_{i}) > p_{i}(s_i, \mu_{i}),
\]

\item $G \Ra G'$ when $G \neq G'$ and for all $i \in [1..n]$ 
\[
\fa s_i \in S_i \setminus S'_i \: \fa \mu_{i} \in {\cal B}_{i} \stackrel{.}{\cap} G \: \te s'_i \in S'_i \:
p_{i}(s'_i, \mu_{i}) > p_{i}(s_i, \mu_{i}).
\]
\end{itemize}

So in these two alternatives we refer to better responses in,
respectively, $G$ and in $G'$ instead of in $H$.  

Clearly $G \Ra G'$ implies $G \myra G'$ which implies $G \leadsto G'$. 
However, the reverse implications do not need to hold.
The following example additionally shows that neither
$\myra$ nor $\Ra$ is
order independent. Moreover, countable applications of each of these
two relations can reduce the initial game to an empty game.

\begin{example} \label{exa:1a}
  Reconsider the two-players game $H$ from Example \ref{exa:1}. Recall
  that the set of strategies for each player in $H$ is the set of
  natural numbers ${\cal N}$ and the payoff to each player is the
  number (strategy) he selected.  Also, we assume the pure belief structure.

Given two subsets $A_1, A_2$ of the set of natural numbers denote by 
$(A_1, A_2)$ the restriction of $H$ in which $A_i$ is the set of strategies of 
player $i$. Clearly for all $k \geq 0$ we have
$H \leadsto (\C{k}, \C{k}) \leadsto \ES_2$ and
$H \myra (\C{k}, \C{k})$ and for no $k \geq 0$ and
$G$ we have
$(\C{k}, \C{k}) \myra G$.

So the relations $\leadsto$ and $\myra$ differ
in the iterations starting at $H$. Moreover, $\myra$ is
not order independent.

Further, for no $k \geq 0$ we have $H \Ra (\C{k}, \C{k})$, so the relations
$\myra$ and $\Ra$ differ, as well.
Let now for $k \geq 0$
\[
(k, \infty) := \C{\ell \in {\cal N} \mid \ell > k},
\]
\[
A_k := \C{0} \cup (k, \infty),
\]
\[
B_k := \C{1} \cup (k, \infty).
\]
Then both
\[
H \Ra (A_1, A_1) \Ra (A_2, A_2) \Ra \ \LL
\]
and 
\[
H \Ra (B_1, B_1) \Ra (B_2, B_2) \Ra \ \LL,
\]
so both
$H \Ra^{\hspace{-1mm} \omega} (\C{0}, \C{0})$
and $H \Ra^{\hspace{-1mm} \omega} (\C{1}, \C{1})$.
But for no $G$ and $k \geq 0$ we have
$(\C{k}, \C{k}) \Ra G$.
This shows that $\Ra$ is not order independent either.

Finally, note that 
\[
H \Ra ((0, \infty), (0, \infty)) \Ra ((1, \infty), (1, \infty)) \Ra \ \LL
\]
so $H \Ra^{\hspace{-1mm} \omega} \ES_2$ and hence
$H \myra^{\hspace{-1mm} \omega} \ES_2$, as well.
In fact, we also have $H \myra \ES_2$.
\HB
\end{example}

Let us define now the counterpart $^{f} \hspace{-1mm} \myra$ of the $^{f} \hspace{-1mm} \leadsto$ reduction
by putting for
a restriction \\ $G := (S_1, \LL, S_n, p_1, \LL, p_n)$ of $H$ and a
restriction $G' := (S'_1, \LL, S'_n, p_1, \LL, p_n)$ of $G$

\begin{itemize}
\item $G \hspace{1mm} ^{f} \hspace{-1mm} \myra G'$ 
when $G \neq G'$ and for all $i \in [1..n]$ 
\[
S'_i = \{ s_i \in S_i \mid \te \mu_{i} \in {\cal B}_{i} \stackrel{.}{\cap} G \: \fa s'_i \in S_i \:
p_{i}(s'_i, \mu_{i}) \leq p_{i}(s_i, \mu_{i})\}.
\]
\end{itemize}

So, unlike in the definition of the $^{f} \hspace{-1mm} \leadsto$ relation, we now refer to 
the best responses in the game $G$.
Note that $G \hspace{1mm} ^{f} \hspace{-1mm} \myra G'$
and $G \myra G''$ implies $G' \sse G''$.
In \cite{Ber84} and \cite{Pea84} the
$^{f} \hspace{-1mm} \myra$ reduction was studied, in each paper for a specific belief structure.

Observe that the corresponding `fast' reduction $^{f} \hspace{-1mm} \Ra$
does not exist. Indeed, in the above example we have 
$H \Ra ((k, \infty), (k, \infty))$ for all $k
\geq 0$.  But $\bigcap_{k = 0}^{\infty} (k, \infty) = \ES$ and $H
  \Ra \ES_2$ does not hold.
So no $G'$ exists such that $H \Ra G'$ and for all $G''$, 
$G \Ra_{\hspace{-1mm}} G''$ implies $G' \sse G''$.

In the game used above we have both $H \hspace{1mm}^{f} \hspace{-1mm}
\leadsto \ES_2$ and $H \hspace{1mm} ^{f} \hspace{-1mm} \myra \ES_2$,
so both fast reductions coincide when started at $H$.  The next
example shows that this is not the case in general.  Moreover, it
demonstrates that for the $\myra$ relation a stronger reduction can be
achieved if non-fast reductions are allowed. So the counterpart of
Theorem \ref{thm:fast} does not hold for the $\myra$ relation.

\begin{example} \label{exa:Bertrand}
Consider a version of Bertrand competition between two firms
in which the marginal costs are 0 and in which
the range of possible prices is the left-open real interval $(0, 100]$.
So in this game $H$ there are two players, each with the set $(0, 100]$ of strategies.
We assume that the demand equals $100 - p$, where $p$ is the lower price and
that the profits are split in case of a tie.
So the payoff functions are defined by:

\[
p_1(s_1, s_2) := \left\{ 
\begin{tabular}{ll}
$s_1 (100 - s_1) $ &  \mbox{if $s_1 < s_2$} \\[2mm]
$\dfrac{s_1 (100 - s_1)}{2} $ &  \mbox{if $s_1 = s_2$} \\[2mm]
0 &  \mbox{if $s_1 > s_2$} 
\end{tabular}
\right . 
\]
\[
p_2(s_1, s_2) := \left\{ 
\begin{tabular}{ll}
$s_2 (100 - s_2) $ &  \mbox{if $s_2 < s_1$} \\[2mm]
$\dfrac{s_2 (100 - s_2)}{2} $ &  \mbox{if $s_1 = s_2$} \\[2mm]
0 &  \mbox{if $s_2 > s_1$} 
\end{tabular}
\right . 
\]

Also, we assume the pure belief structure.  Below we identify the
restrictions of $H$ with the pairs of the strategy sets of the
players.

Since $s_1 = 50$ maximizes the value of $s_1 (100 - s_1)$ in the
interval $(0, 100]$, the strategy 50 is the unique best response to
any strategy $s_2 > 50$ of the second player.  Further, no strategy is
a best response to a strategy $s_2 \leq 50$.  By symmetry the same holds
for the strategies of the second player.  So $H \hspace{1mm} ^{f} \hspace{-1mm} \leadsto
(\C{50}, \C{50})$.
Next, $s_1 = 49$ is a better response in $H$ to $s_2 = 50$ than $s_1 =
50$ and symmetrically for the second player. So $(\C{50}, \C{50}) \hspace{1mm} ^{f}
\hspace{-1mm} \leadsto \ES_2$.

We also have $H \hspace{1mm} ^{f} \hspace{-1mm} \myra (\C{50},
\C{50})$.  But $s_1 = 50$ is a best response in $(\C{50}, \C{50})$ to
$s_2 = 50$ and symmetrically for the second player. So for no
restriction $G$ of $H$ we have $(\C{50}, \C{50}) \myra G$ or $(\C{50},
\C{50}) \hspace{1mm} ^{f} \hspace{-1mm} \myra G$.  However, we also
have $H \myra ((0, 50], (0, 50]) \myra \ES_2$.  So $H$ can be reduced
to the empty game using the $\myra$ reduction but only if non-fast
reductions are allowed.

Finally, note that also $H \Ra ((0, 50], (0, 50])$ holds.  Let
$(r_i)_{i < \omega}$ be a strictly descending sequence of real numbers
starting with $r_0 = 50$ and converging to 0.  It is easy to see that
for $i \geq 0$ we then have $((0, r_i], (0, r_i]) \Ra ((0, r_{i+1}], (0, r_{i+1}])$, so
$H \Ra^{\hspace{-1mm} \omega} \hspace{1mm} \ES_2$.
\HB
\end{example}

To analyze the situation when the three considered reduction relations coincide
we introduce the following property:
\begin{description}

\item[B] For all restrictions $G$ of $H$ and all beliefs $\mu_i \in {\cal B}_{i} \stackrel{.}{\cap} G$
a best response to $\mu_i$ in $G$ exists.
\end{description}

For the finite games property \textbf{B} obviously holds.  However, it
can fail for infinite games. For instance, it does not hold in the
game considered in Examples \ref{exa:1} and \ref{exa:1a} since in this
game no strategy is a best response to a strategy of the opponent.

In the presence of property \textbf{B} the reductions $\myra$ and 
$\Ra$ are equivalent.

\begin{lemma}[Equivalence] \label{lem:equb} 
Consider a game $H$ and a belief
structure $({\cal B}, \stackrel{.}{\cap})$ in $H$.
Assume property \textbf{B}.
The relations $\myra$ and $\Ra$ coincide on the set of restrictions of $H$.

\end{lemma}
\Proof
Clearly if $G \Ra G'$, then
$G \myra G'$.
To prove the converse let $G := (S_1, \LL, S_n, p_1, \LL, p_n)$, $G' := (S'_1, \LL, S'_n,
p_1, \LL, p_n)$ and
${\cal B} := ({\cal B}_{1}, \LL,  {\cal B}_{n})$.

Suppose $G \myra G'$. Take an arbitrary
$s_i \in S_i \setminus S'_i$ 
and an arbitrary $\mu_i \in {\cal B}_{i} \stackrel{.}{\cap} G$. 
By property \textbf{B} some
$s'_i \in S_i$ is a best response to $\mu_i$ in $G$.  By
definition this $s'_i$ is not eliminated in the step $G
\myra G'$, i.e., $s'_i \in S'_i$.  
So $s_i$ is not a best response to $\mu_i$ in $G'$.
This proves $G \Ra G'$.
\HB 
\VV

However, the situation changes when we consider the $\leadsto$ relation. 
We noted already that $G \myra G'$
implies $G \leadsto G'$. But the converse does not need to hold,
even if property \textbf{B} holds.

\begin{example}
Suppose that $H$ equals
\begin{center}
\begin{game}{3}{2}
      & $L$    & $R$ \\
$T$   &$2,0$   &$2,0$\\
$M$   &$0,0$   &$1,0$ \\
$B$   &$1,0$   &$0,0$
\end{game}
\end{center}
$G$ is
\begin{center}
\begin{game}{2}{2}
      & $L$    & $R$ \\
$M$   &$0,0$   &$1,0$ \\
$B$   &$1,0$   &$0,0$
\end{game}
\end{center}
and $G'$ is
\begin{center}
\begin{game}{1}{2}
      & $L$    & $R$ \\
$M$   &$1,0$   &$1,0$ \\
\end{game}
\end{center}

Further, assume the pure belief structure.
Property \textbf{B} holds since the game $H$ is finite.

Since the strategy $B$ is never a best response to a strategy of the opponent in the game $H$,
we have $G \leadsto G'$ but $G \myra G'$ does not hold
since $B$ is a best response to $L$ in the game $G$.
\HB
\end{example}

On the other hand, in the presence of properties \textbf{A} and \textbf{B}, 
iterated applications of the $\leadsto$
reduction started in $H$ do yield the same outcome as
the iterated applications of  $\myra$
or of $\Ra$. 
Indeed, the following holds. 

\begin{lemma}[Equivalence] \label{lem:equ2}
Consider a game $H$ and a belief
structure $({\cal B}, \stackrel{.}{\cap})$ in $H$.
Assume properties \textbf{A} and \textbf{B}.
For all restrictions $G$ of $H$,  $H \leadsto^{\gamma} G$ iff
$H \myra^{\hspace{-1mm} \gamma} \hspace{1mm} G$.
\end{lemma}

\Proof
Since $G' \myra G''$ implies
$G' \leadsto G''$, for all $\gamma$
$H \myra^{\hspace{-1mm} \gamma} \hspace{1mm} G$ implies
$H \leadsto^{\gamma} G$.

To prove the converse we proceed by transfinite induction.
Assume that $H \leadsto^{\gamma} G$.
By definition an appropriate transfinite sequence of restrictions
$(G_{\alpha})_{\alpha \leq \gamma}$ of $H$ with $H = G_{0}$ and $G_{\gamma} =
G$ exists ensuring that $H \leadsto^{\gamma} G$.

Suppose the claim of the lemma holds for all $\beta < \gamma$.
\II

\NI
\emph{Case 1}. $\gamma$ is a successor ordinal, say $\gamma = \beta +1$.

Then $H \leadsto^{\beta} G_{\beta}$ and 
$G_{\beta} \leadsto G$. Suppose that
${\cal B} := ({\cal B}_{1}, \LL,  {\cal B}_{n})$,
$
H := (T_1, \LL, T_n, p_1, \LL, p_n),
$
$
G_{\beta} := (S_1, \LL, S_n, p_1, \LL, p_n)
$
and 
$
G := (S'_1, \LL, S'_n, p_1, \LL, p_n).
$

Consider an arbitrary $s_i \in S_i \setminus S'_i$ and 
an arbitrary $\mu_{i} \in
{\cal B}_{i} \stackrel{.}{\cap} G_{\beta}$ such that $s_i$ is not a best
response in $H$ to $\mu_{i}$.  By property \textbf{A}
\begin{equation}
  \label{eq:mu}
\mbox{$\mu_{i} \in {\cal B}_{i} \stackrel{.}{\cap} G_{\alpha}$ for all $\alpha \leq \beta$.}
\end{equation}

By property \textbf{B} a best response
$s'_i$ to $\mu_i$ in $H$ exists.  Then $p_i(s'_i, \mu_i) > p_i(s_i, \mu_i)$ and
$p_i(s'_i, \mu_i) \geq p_i(s''_i, \mu_i)$ for all $s''_i \in T_i$.
By the latter inequality and (\ref{eq:mu}) $s'_i$ 
is not removed in any $\leadsto$ step leading from $H$ to $G_{\beta}$. So $s'_i \in S_i$
and by the former (strict) inequality 
$s_i$ is not a best response to $\mu_i$ in $G_{\beta}$.
This proves $G_{\beta} \myra G$.
But by the induction hypothesis
$H \myra^{\beta} G_{\beta}$, so $H \myra^{\gamma} G$.
\II

\NI
\emph{Case 2}. $\gamma$ is a limit ordinal.

By the induction hypothesis for all $\beta < \gamma$ we have
$H \leadsto^{\beta} G_{\beta}$ iff
$H \myra^{\hspace{-1mm} \beta} G_{\beta}$, so
by definition 
$H \leadsto^{\gamma} G$ iff
$H \myra^{\hspace{-1mm} \gamma} G$.
\HB
\VV

This allows us to establish an order independence result for the 
$\myra$ and $\Ra$
relations.

\begin{theorem}[Order Independence] \label{thm:order1}
  Consider a game $H$ and a belief
structure $({\cal B}, \stackrel{.}{\cap})$ in $H$.
  Assume properties \textbf{A} and \textbf{B}.  

\begin{enumerate} \smallromani
  \item All maximal
  sequences of the $\myra$, $\Ra$ and
$\leadsto$ reductions yield the same outcome $G$.

  \item 
This restriction $G$ satisfies the following property:

each strategy $s_i$ of player $i$ in $G$ is a best response in $G$ (note
this reference to $G$ and not $H$) to a belief in ${\cal B}_{i} \stackrel{.}{\cap} G$.  
  \end{enumerate}

\end{theorem}
\Proof

\NI
$(i)$
By the Order Independence Theorem \ref{thm:order} and 

\II

\noindent
$(ii)$ By $(i)$ for no $G'$ we have $G \myra G'$,
which proves the claim.
\HB

%




\section{Beliefs as joint pure strategies of the opponents}
\label{sec:opponents}

So far we established results for arbitrary belief structures that
satisfy properties \textbf{A} and \textbf{B}.  In this section we
analyze what additional properties hold
for the case of pure belief structures.  So given a game $H :=
(T_1, \LL, T_n, p_1, \LL, p_n)$ we assume ${\cal B}_{i} := T_{-i}$ for
$i \in [1..n]$ and for a restriction $G := (S_1, \LL, S_n, p_1, \LL,
p_n)$ of $H$ we assume $T_{-i} \stackrel{.}{\cap} G := S_{-i}$.

Clearly, property \textbf{A} then holds.  By the 
Order Independence Theorem \ref{thm:order}, the outcome of each
maximal sequence of the $\leadsto$ reductions is unique. We noted
already that this outcome can be an empty game.  On the other hand,
if the initial game has a Nash equilibrium, then
this unique outcome cannot be a degenerate restriction.
Indeed, the following result holds.

\begin{theorem} \label{thm:nonempty}
  Consider a game $H :=  (T_1,\LL, T_n, p_1, \LL, p_n)$. 
Suppose that $H \leadsto^{\gamma} G$ for some $\gamma$.

  \begin{enumerate} \smallromani
  \item If $s$ is a Nash equilibrium of $H$, then it is a Nash
    equilibrium of $G$.  Consequently, if $G$ is empty, then $H$ has
    no Nash equilibrium.

  \item Suppose that for each $s_{-i} \in T_{-i}$ a best response to $s_{-i}$ in $H$ exists.
If $s$ is a Nash equilibrium of $G$, then it is a Nash equilibrium of $H$.
  \end{enumerate}

\end{theorem}

\Proof 

\NI
$(i)$ 
Let $G'$ be the unique outcome of a maximal sequence of the $\leadsto$ 
  reductions that starts with $H \leadsto^{\gamma} G$.
By definition $(s_1, \LL, s_n)$ is a Nash equilibrium of $H$ iff
each $s_i$ is a best response to $s_{-i}$ iff (by the choice of ${\cal
  B}$) $(\C{s_1}, \LL, \C{s_n})$ is ${\cal B}$-closed.  Hence, by the
Order Independence Theorem \ref{thm:order}, each Nash equilibrium $s$ of
$H$ is present in $G'$ and hence in $G$. But $G$ is a restriction of $H$, 
so $s$ is also a Nash equilibrium of $G$. 
\II

\noindent
$(ii)$
Suppose $s$ is not a Nash equilibrium of $H$. Then 
some $s_i$ is not a best response to $s_{-i}$ in $H$.
By assumption a best response  $s'_i$ to $s_{-i}$ in $H$ exists.
Then 
\[
p_{i}(s'_i, s_{-i}) > p_{i}(s).
\]
The strategy $s'_i$ is not eliminated in any $\leadsto$ step leading from $H$ to $G$,
since $s_{-i}$ is a joint strategy of the opponents of player $i$ in all games
in the considered maximal sequence. So $s'_i$ is a strategy of player $i$ in $G$,
which contradicts the fact that $s$ is a Nash equilibrium of $G$.
\HB 
\VV

The above result applies to all three reduction relations since 
$H \myra^{\hspace{-1mm} \gamma} G$ implies $H \leadsto^{\gamma} G$ and $H
\Ra^{\hspace{-1mm} \gamma} G$ implies $H \leadsto^{\gamma} G$.

The assumption used in $(ii)$ is implied by property \textbf{B}.
A natural situation when property \textbf{B} holds is the
following.  We call a game $H := (T_1, \LL, T_n, p_1, \LL, p_n)$
\oldbfe{compact} if the strategy sets are non-empty compact subsets of
a complete metric space and \oldbfe{own-uppersemicontinuous} if each
payoff function $p_i$ is uppersemicontinuous in the $i$th
argument.\footnote{Recall that $p_i$ is \oldbfe{uppersemicontinuous in
    the} $i$\oldbfe{th argument} if the set $\C{s'_i \in T_i \mid
    p_i(s'_i, s_{-i}) \geq r}$ is closed for all $r \in {\cal R}$ and
  all $s_{-i} \in T_{-i}$.}  

As explained in \cite{DS02} (see the proof of Lemma on page 2012) for
such games property \textbf{B} holds by virtue of a standard result
from topology.  Consequently, by the Order Independence Theorem \ref{thm:order1}, the
order independence for the $\myra$, $\Ra$ and $\leadsto$ reduction
relations holds.  Let us also mention that for this class of
games \cite{DS02} established order independence of the iterated
elimination of strictly dominated strategies.

If we impose a stronger condition on the payoff functions, namely that
each of them is continuous, then we are within the framework
considered in \cite{Ber84}. As shown in this paper if at each stage
the $^{f} \hspace{-1mm} \myra$ reduction is applied,
the final (unique) outcome is a non-degenerate restriction and is
reached after at most $\omega$ steps.  
This allows us to draw the following corollary to
the Order Independence Theorems \ref{thm:order} and \ref{thm:order1}.

\begin{corollary} \label{cor:ber84}
  Consider a compact game $H$ with continuous payoff functions.
  All maximal sequences of the $\leadsto$ (or $\myra$ or $\Ra$)
reductions starting in $H$ yield the same outcome which is a non-degenerate
restriction of $H$.  
\HB
\end{corollary}

If at each stage only some strategies that are NBR are
removed, transfinite reduction sequences of length $> \omega$ are
possible.  In Section \ref{sec:reductions} we already noted that in some
games such transfinite sequences are unavoidable.

Let us mention here that under the same assumptions about the game $H$
\cite{Amb94} showed the analogue of the Equivalence Lemma
\ref{lem:equ2} for the `fast' counterparts $^{f} \hspace{-1mm}
\leadsto$ and $^{f} \hspace{-1mm} \myra$ of the reduction relations
$\leadsto$ and $\myra$, for the limited case of two-person games and
beliefs equal to the strategies of the opponent.
\footnote{To be precise, in his definition the `fast' reductions are defined
by considering the reduction for each player in succession and not in parallel.}
By the
abovementioned result of \cite{Ber84} the corresponding iterations of
these two reduction relations reach the final outcome after at most
$\omega$ steps.

Recall now that a simple strengthening of the assumptions of
\cite{Ber84} leads to a framework in which existence of a (pure) Nash
equilibrium is ensured.  Namely, assume that strategy sets are
non-empty compact convex subsets of a complete metric space and each
payoff function $p_i$ is continuous and quasi-concave in the $i$th
argument.\footnote{Recall that $p_i$ is \oldbfe{quasi-concave in the}
  $i$\oldbfe{th argument} if the set $\C{s'_i \in T_i \mid p_i(s'_i,
    s_{-i}) \geq p_i(s)}$ is convex for all $s \in T$.}  By a theorem
of \cite{Deb52}, \cite{Fan52} and \cite{Gli52} under these assumptions a Nash
equilibrium exists.

Natural examples of games satisfying these assumptions
are \oldbfe{mixed extensions} of
finite games, i.e., games in which the players' strategies are their
mixed strategies in a finite game $H$ and the payoff functions are the
canonic extensions of the payoffs in $H$ to the joint mixed
strategies.

Let us modify now the definition of the narrowing operation $\stackrel{.}{\cap}$
by putting for a mixed extension
$H := (T_1, \LL, T_n, p_1, \LL, p_n)$ and its restriction $G := (S_1, \LL, S_n, p_1, \LL,
p_n)$

\begin{equation}
  \label{eq:convex}
T_{-i} \stackrel{.}{\cap} G := \Pi_{j \neq i} \overline{S_{j}},  
\end{equation}
where for a set $M_j$ of mixed strategies of player $j$ $\overline{M_{j}}$ denotes its convex hull.
Then, as before, properties \textbf{A} and \textbf{B} hold.

This situation corresponds to the setup of \cite{Pea84} in which at
each stage \emph{all} mixed strategies that are NBR are
deleted and $\stackrel{.}{\cap}$ is defined by ($\ref{eq:convex}$).
\cite{Pea84} proved that this iterative process based on 
the $^{f} \hspace{-1mm} \myra$ reduction terminates after finitely
many steps and yields a non-degenerate restriction.
So we get another corollary to
the Order Independence Theorem \ref{thm:order} and the Equivalence
Lemmata \ref{lem:equb} and \ref{lem:equ2}.

\begin{corollary}
  Let $H := (T_1, \LL, T_n, p_1, \LL, p_n)$ be a mixed extension of a finite game.
  Suppose that $\stackrel{.}{\cap}$
is defined by ($\ref{eq:convex}$). 
  Then all maximal sequence of the $\leadsto$ (or
  $\myra$ or $\Ra$) reductions
  yield the same outcome which is a non-degenerate restriction of $H$.  
\HB
\end{corollary}

The same outcome is obtained when at each stage only some mixed
strategies that are NBR are deleted.  In this case the
iteration process can be infinite, possibly continuing beyond
$\omega$.

\section{Finite games}
\label{sec:finite}

Finally, we consider the case of \oldbfe{finite} games,  i.e., ones in which all strategy sets are finite.
Given a finite non-empty set $A$ we denote by $\Delta A$ the set of probability
distributions over $A$.  

Consider a finite game $H := (T_1, \LL, T_n, p_1, \LL, p_n)$.  In what 
follows by a \oldbfe{belief of player $i$} in the game $H$ we mean a
probability distribution over the set of joint strategies of his
opponents.  So $\Delta T_{-i}$ is the set of beliefs.  The payoff
functions $p_i$ are modified to the expected payoff functions in the
standard way by putting for $\mu_i \in \Delta T_{-i}$:
\[
p_i(s_i, \mu_i) := \sum_{s_{-i} \in T_{-i}} \mu_i(s_{-i}) \cdot p_i(s_i, s_{-i}).
\]

We noted already that for the finite games property \textbf{B} obviously holds. For further
considerations we need the following property:

\begin{description}
\item[C] For all non-degenerate restrictions $G$ of $H$, ${\cal B}_{i} \stackrel{.}{\cap} G \neq \ES$.
\end{description}

\begin{note} \label{cor:finite}
  Consider a finite game $H:= (T_1, \LL, T_n, p_1, \LL, p_n)$ and a
  belief structure $({\cal B}, \stackrel{.}{\cap})$ in $H$. Assume
  properties \textbf{A}-\textbf{C}.  Then a non-degenerate ${\cal
    B}$-closed restriction of $H$ exists.
\end{note}
\Proof 
Keep applying the $\leadsto$ reduction starting
with the original game $H$. Since now only finite sequences of
$\leadsto$ reductions exist, this iteration process stops
after finitely many steps.  By the Equivalence Lemma \ref{lem:equ2}
its outcome coincides with the repeated application of the $\myra$ reduction. 

But by definition, in the presence of properties \textbf{A}-\textbf{C}, if
$G$ is a non-degenerate restriction of $H$ and
$G \myra G'$, then $G'$ is
non-degenerate, as well.  So in this iteration process only
non-degenerate restrictions are produced.  
\HB 
\VV

Three successively larger sets of beliefs are of interest:

\begin{itemize}
\item ${\cal B}_{i} = T_{-i}$ for $i \in [1..n]$.

Then beliefs are joint pure strategies of the opponents.

\item ${\cal B}_{i} = \Pi_{j \neq i} \Delta  T_{j}$ for $i \in [1..n]$.

Then beliefs are joint mixed strategies of the opponents.

\item ${\cal B}_{i} = \Delta T_{-i}$ for $i \in [1..n]$.

Then beliefs are probability distributions over the set of joint pure strategies of the opponents.
\end{itemize}

These sets of beliefs are increasingly larger in the sense that we can
identify $T_{-i}$ with the subset of $\Pi_{j \neq i} \Delta T_{j}$
consisting of the joint pure strategies and in turn $\Pi_{j \neq i} \Delta
T_{j}$ with the subset of $\Delta T_{-i}$ consisting of the so-called
\emph{uncorrelated} beliefs.

For ${\cal B}_{i} \sse \Delta T_{-i}$ and $G := (S_1, \LL, S_n, p_1, \LL,
p_n)$ we define then
\begin{equation}
  \label{eq:beliefs}
{\cal B}_{i} \stackrel{.}{\cap} G := \{\mu_i \in {\cal B}_{i} \mid 
\mbox{$\mu_i(s_{-i}) = 0$ for  $s_{-i} \in T_{-i} \setminus S_{-i}$\}.}
\end{equation}

In particular, in view of the above identifications,
\[
T_{-i} \stackrel{.}{\cap} G = S_{-i},
\]
\[
\Pi_{j \neq i} \Delta  T_{j} \stackrel{.}{\cap} G = \Pi_{j \neq i} \Delta  S_{j},
\]
and
\[
\Delta T_{-i} \stackrel{.}{\cap} G = \Delta S_{-i}.
\]

Consider now properties \textbf{A} and \textbf{C}.  Property
\textbf{A} obviously holds.  In turn, 
property \textbf{C} holds if
$T_{-i} \sse {\cal B}_{i}$ for all $i \in [1..n]$.  

Summarizing, in view of Note \ref{cor:finite}
we get the following corollary to the 
Order Independence Theorem \ref{thm:order} and 
the Equivalence Lemmata \ref{lem:equb} and \ref{lem:equ2}.

\begin{corollary} \label{cor:iter}
  Consider a finite game $H := (T_1, \LL, T_n, p_1, \LL, p_n)$ and
  suppose that $T_{-i} \sse {\cal B}_{i}$ for all $i \in [1..n]$
and that $\stackrel{.}{\cap}$ is defined by (\ref{eq:beliefs}).
  Then all maximal sequences of the $\leadsto$ (or
  $\myra$ or $\Ra$)
  reductions yield the same outcome which is a non-degenerate
  restriction of $H$.  \HB
\end{corollary}

In particular, each set ${\cal B}_{i}$ can be instantiated to any of
the three sets of beliefs listed above.  However, the assumption that
$T_{-i} \sse {\cal B}_{i}$ for all $i \in [1..n]$ excludes systems of
beliefs ${\cal B} := ({\cal B}_{1}, \LL, {\cal B}_{n})$ that consist
of the joint totally mixed strategy of the opponents.  Recall that a mixed
strategy is called \oldbfe{totally mixed} if it assigns a positive
probability to each pure strategy.  Indeed, any element $s_{-i} \in
T_{-i}$ is a sequence of pure strategies of the opponents of player
$i$ and each such pure strategy $s_j$ is identified with a mixed strategy
that puts all weight on $s_j$ (and hence weight zero on
other pure strategies).  So no element of $s_{-i}$ from $T_{-i}$ can be identified
with a totally mixed strategy.

Observe also that if each ${\cal B}_{i}$ is the set of joint totally
mixed strategy of the opponents, then for each proper restriction $G$
of $H$ the sets ${\cal B}_{i} \stackrel{.}{\cap} G$ are all empty. So
${\cal B}_{i} \stackrel{.}{\cap} G$ does not model the set of joint
totally mixed strategies of the opponents of player $i$ in the game
$G$.

The systems of beliefs involving totally mixed strategies were
studied in several papers, starting with \cite{Pea84}, where a best
response to a belief formed by a joint totally mixed strategy of the
opponents is called a \oldbfe{cautious response}. A number of
modifications of the notion of rationalizability rely on a specific
use of totally mixed strategies, see, e.g., \cite{HV00} where the
notion of weak perfect rationalizability is studied.  

Corollary \ref{cor:iter}
does not apply to the iterated elimination procedures based on
such systems of beliefs. This is not surprising, since 
as shown in \cite{Pea84} a strategy is weakly dominated iff it is never a
cautious response, and weak dominance is order dependent.  In turn,
the elimination procedure discussed in \cite{HV00} is shown to be
equivalent to the \cite{DF90} elimination procedure
which consists of one round of elimination of all weakly dominated
strategies followed by the iterated elimination all strictly dominated
strategies.

\section{Concluding remarks}
\label{sec:conc}

We studied in this paper the problem of order independence for
rationalizability in strategic games. To this end we relaxed the
requirement that at each stage all strategies that are never best
responses are eliminated. This brought us to a study of three natural
reduction relations.

The iterated elimination of NBR is supposed to
model reasoning of a rational player, so we should reflect on the
consequences of the obtained results. First, we noted that in some
games the transfinite iterations can be unavoidable.  This difficulty
was already discussed in \cite{Lip94} who concluded that finite order
mutual knowledge may be insufficient as a characterization of common
knowledge.

Next, we noted that in the natural situation when beliefs are the
joint pure strategies of the opponents empty games can be generated
using each of the reduction relations $\leadsto$,
$\myra$ and $\Ra$.
We could interpret such a situation as a statement that in the initial
game no player has a meaningful strategy to play.  Note that Theorem
\ref{thm:nonempty} allows us to conclude that the initial game has
then no Nash equilibrium.

Another issue is which of the three reduction relations is the `right'
one.  The first one considered, $\leadsto$, is the
\emph{strongest} in the sense that its iterated applications achieve
the strongest reduction. It is order independent under a very weak
assumption \textbf{A} that captures the idea of a `well-behaving'
belief structure.

However, its definition refers to the
strategies of the initial game $H$ which at the moment of reference
may already have been discarded. This point can be illustrated using the
Bertrand competition game of Example \ref{exa:Bertrand}.  We concluded
there that $(\C{50}, \C{50}) \leadsto \ES_2$ because $s_1
= 49$ is a better response of the first player in $H$ to $s_2 = 50$
than $s_1 = 50$ and symmetrically for the second player. However, the
strategy $s_1 = 49$ is already discarded at the moment the game
$(\C{50}, \C{50})$ is considered, so ---one might argue--- it should
not be used to discard another strategy. If one accepts this
viewpoint, then one endorses $\myra$ as the
right reduction. 
This reduction relation is not order independent under
assumption \textbf{A} but is order independent once we add
assumption \textbf{B} stating that for each belief $\mu_i$ in a
restriction $G$ of the original game a best response to $\mu_i$ in $G$
exists. 

Finally, we can view the $\Ra$ reduction as a `conservative' variant
of $\myra$ in which one insists that the `witnesses' used to discard
the strategies should not themselves be discarded (in the same round).
In the case of the iterated elimination of strictly dominated
strategies the corresponding reduction relation was studied in
\cite{GKZ90} and \cite{DS02}.

Note that the difficulty of chosing the right reduction relation does
not arise in \cite{Ber84} and \cite{Pea84} since for the class of the games there
studied properties \textbf{A} and \textbf{B} hold and consequently the
Equivalence Lemmata \ref{lem:equb} and \ref{lem:equ2} can be applied.

In the previous two sections we established order independence for all
three reduction relations $\leadsto, \myra$ and $\Ra$ for the same
classes of games for which order independence of the iterated
elimination of strictly dominated strategies (SDS) holds.  It was
already indicated in \cite{Pea84} that the iterated elimination of NBR
yields a stronger reduction than the iterated elimination of SDS.  The
Bertrand competition game of Example \ref{exa:Bertrand} provides an
illuminating example of this phenomenon.  In this game all three
reduction relations $\leadsto, \myra$ and $\Ra$ allow us to reduce the
initial game to an empty one. However, in this game no strategy
strictly dominates another one.  Indeed, for any $s_1$ and $s'_1$ such
that $0 < s_1 < s'_1 \leq 100$ we have $p_1(s_1, s_2) = p_1(s'_1, s_2)
= 0$ for all $s_2$ such that $0 < s_2 < s_1$ and analogously for the
second player.  So no strategy can be eliminated on the account of
strict dominance.

This advantage of each variant of the iterated elimination of NBR
over the iterated elimination of SDS disappears if we provide each player with
the strategy 0.  Then each strategy is a best response to the strategy
0 of the opponent, since all of them yield the same payoff, 0. So no
strategy can be eliminated and all four elimination methods yield no
reduction, while the resulting game has a unique Nash equilibrium,
namely $(0,0)$.

Let us conclude with an example when two variants of the iterated
elimination of NBR allow one to identify the unique Nash equilibrium,
while the iterated elimination of SDS yields no reduction.

\begin{example}
Consider Hotelling location game in which two sellers choose a location in the 
open real interval $(0, 100)$.
So in this game $H$ there are two players, each with the set $(0, 100)$ of strategies.
The payoff functions $p_i$ ($i = 1, 2$) are defined by:

\[
p_i(s_i, s_{3-i}) := \left\{ 
\begin{tabular}{ll}
$s_i + \dfrac{s_{3-i} - s_i}{2}$ &  \mbox{if $s_i < s_{3-i}$} \\[2mm]
$100 - s_i + \dfrac{s_i - s_{3-i}}{2}$ &  \mbox{if $s_i > s_{3-i}$} \\[2mm]
$50$ &  \mbox{if $s_i = s_{3-i}$} 
\end{tabular}
\right . 
\]

First note that no strategy strictly dominates another one.  Indeed,
for any $s_1$ and $s'_1$ such that $0 < s_1 < s'_1 < 100$
we have $p_1(s_1, s_2) < p_1(s'_1, s_2)$ for all $s_2$ such that
$s'_1 < s_2 < 100$ and $p_1(s_1, s_2) > p_1(s'_1, s_2) = 0$ for all
$s_2$ such that $0 < s_2 < s_1$.
A symmetric reasoning holds for the second player.

Next, we consider the reduction relations $\leadsto$,
$\myra$ and $\Ra$ defined as in Section \ref{sec:opponents}.
Note that no strategy $s_1 \in (0, 100) \setminus \C{50}$ is a
best response in $H$ to a strategy $s_2 \in (0, 100)$. Indeed, if $s_1
\neq s_2$, then we have $p_1(s_1, s_2) < p_1(s'_1, s_2)$ for all
$s'_1$ such that $s'_1 \in (min(s_1, s_2), max(s_1, s_2)$.
And if $s_1 = s_2$, then by assumption $s_1 \neq 50$ and
we have then $p_1(s_1, s_2) = 50 < p_1(50, s_2)$.
A symmetric reasoning holds for the second player.

So both $H \leadsto (\C{50}, \C{50})$ 
and $H \myra_{\hspace{-1mm}} (\C{50}, \C{50})$ 
and $(\C{50}, \C{50})$ 
is Nash equilibrium of $H$.
Also note that $\Ra_{\hspace{-1mm}}$ yields no reduction here.
\HB  
\end{example}

\section*{Acknowledgements}

We thank one of the referees for useful comments and for drawing our
attention to the paper \cite{Amb94}.



\normalsize

\begin{thebibliography}{1999}

\normalsize

\bibitem[Ambroszkiewicz:\nameindex{Ambroszkiewicz, S.}:1994]{Amb94}
{\sc S.~Ambroszkiewicz\nameindex{Ambroszkiewicz, S.}}, Knowledge and best
  responses in games, {\em Annals of Operations Research}, 51, pp.~63--71.

\bibitem[Bernheim:\nameindex{Bernheim, B.~D.}:1984]{Ber84}
{\sc B.~D. Bernheim\nameindex{Bernheim, B.~D.}}, Rationalizable strategic
  behavior, {\em Econometrica}, 52, pp.~1007--1028.

\bibitem[Debreu:\nameindex{Debreu, G.}:1952]{Deb52}
{\sc G.~Debreu\nameindex{Debreu, G.}}, A social equilibrium existence theorem,
  {\em Proceedings of the National Academy of Sciences}, 38, pp.~886--893.

\bibitem[Dekel and Fudenberg:\nameindex{Dekel, E.}\nameindex{Fudenberg,
  D.}:1990]{DF90}
{\sc E.~Dekel\nameindex{Dekel, E.} and D.~Fudenberg\nameindex{Fudenberg, D.}},
  Rational behavior and payoff uncertainty, {\em Journal Of Economic Theory},
  pp.~1391--1402.

\bibitem[Dufwenberg and Stegeman:\nameindex{Dufwenberg, M.}\nameindex{Stegeman,
  M.}:2002]{DS02}
{\sc M.~Dufwenberg\nameindex{Dufwenberg, M.} and
  M.~Stegeman\nameindex{Stegeman, M.}}, Existence and uniqueness of maximal
  reductions under iterated strict dominance, {\em Econometrica}, 70,
  pp.~2007--2023.

\bibitem[Fan:\nameindex{Fan, K.}:1952]{Fan52}
{\sc K.~Fan\nameindex{Fan, K.}}, Fixed point and minimax theorems in locally
  convex topological linear spaces, {\em Proceedings of the National Academy of
  Sciences}, 38, pp.~121--126.

\bibitem[Gilboa, Kalai and Zemel:\nameindex{Gilboa, I.}\nameindex{Kalai,
  E.}\nameindex{Zemel, E.}:1990]{GKZ90}
{\sc I.~Gilboa\nameindex{Gilboa, I.}, E.~Kalai\nameindex{Kalai, E.}, and
  E.~Zemel\nameindex{Zemel, E.}}, On the order of eliminating dominated
  strategies, {\em Operation Research Letters}, 9, pp.~85--89.

\bibitem[Glicksberg:\nameindex{Glicksberg, I.~L.}:1952]{Gli52}
{\sc I.~L. Glicksberg\nameindex{Glicksberg, I.~L.}}, A further generalization
  of the {Kakutani} fixed point theorem, with application to {Nash} equilibrium
  points, {\em Proceedings of the American Mathematical Society}, 3,
  pp.~170--174.

\bibitem[Herings and Vannetelbosch:\nameindex{Herings,
  P.~J.-J.}\nameindex{Vannetelbosch, V.~J.}:2000]{HV00}
{\sc P.~J.-J. Herings\nameindex{Herings, P.~J.-J.} and V.~J.
  Vannetelbosch\nameindex{Vannetelbosch, V.~J.}}, The equivalence of the
  {Dekel-Fudenberg} iterative procedure and weakly perfect rationalizability,
  {\em Economic Theory}, pp.~677--687.

\bibitem[Lipman:\nameindex{Lipman, B.~L.}:1994]{Lip94}
{\sc B.~L. Lipman\nameindex{Lipman, B.~L.}}, A note on the implications of
  common knowledge of rationality, {\em Games and Economic Behavior}, 6,
  pp.~114--129.

\bibitem[Osborne and Rubinstein:\nameindex{Osborne,
  M.~J.}\nameindex{Rubinstein, A.}:1994]{OR94}
{\sc M.~J. Osborne\nameindex{Osborne, M.~J.} and
  A.~Rubinstein\nameindex{Rubinstein, A.}}, {\em A Course in Game Theory}, The
  {MIT} Press, Cambridge, Massachusetts.

\bibitem[Pearce:\nameindex{Pearce, D.~G.}:1984]{Pea84}
{\sc D.~G. Pearce\nameindex{Pearce, D.~G.}}, Rationalizable strategic behavior
  and the problem of perfection, {\em Econometrica}, 52, pp.~1029--1050.

\bibitem[Stegeman:\nameindex{Stegeman, M.}:1990]{Ste90}
{\sc M.~Stegeman\nameindex{Stegeman, M.}}, {\em Deleting strictly eliminating
  dominated strategies}.
\newblock Working Paper 1990/6, Department of Economics, University of North
  Carolina.

\end{thebibliography}

\end{document}